# Adsorption of polyelectrolytes onto the oppositely charged surface of tubular J-aggregates of a cyanine dye

Omar Al-Khatib,[†,‡] Christoph Böttcher,[§] Hans von Berlepsch,[§] Katherine Herman,[†,‡] Sebastian Schön,[††] Jürgen P. Rabe,[†,‡] Stefan Kirstein [†*]

[†] Institut für Physik, Humboldt-Universität zu Berlin

[‡] IRIS Adlershof, Humboldt-Universität zu Berlin

[§] Research Center of Electron Microscopy, Freie Universität Berlin

[††] Institut für Chemie, Technische Universität Berlin

AUTHOR EMAIL: kirstein@physik.hu-berlin.de

**RECEIVED DATE**

TITLE RUNNING HEAD: Wrapping of tubular J-aggregates

CORRESPONDING AUTHOR: Newtonstr. 15, 12489 Berlin, FAX: +49-30-2093-7632, Phone: +49-30-2093-7821

ABSTRACT: The adsorption of three different polycations at the negatively charged surface of tubular J-aggregates of the amphiphilic cyanine dye 3,3'-bis(2-sulfopropyl)-5,5',6,6'-tetrachloro-1,1'-dioctylbenzimidacarbocyanine (C8S3) is investigated by means of cryogenic electron microscopy and optical absorption spectroscopy. All three polycations could be adsorbed at the aggregates without flocculation or precipitation when added in molar amounts of monomers sufficiently smaller than of the dye molecules. Preferably the aggregates are either coated by the polycations in total or left completely uncoated. For the coated aggregates, the adsorption leads to charge reversal of the aggregate surface as supported by zeta-potential measurements. The morphology of the coating differs significantly for the three polycations: The branched polycation polyethlyenimine (PEI) attaches to the tubular aggregate by hit-and-stick adsorption of the coiled state in solution forming irregular clot-like coatings; the flexible and weakly cationic poly(allylamine hydrochloride) (PAH) forms a more homogeneous coating but destroys the integrity of the dye aggregate; the more hydrophobic and strong polycation poly(diallyldimethylammonium chloride) (PDADMAC) forms a thin and homogeneous layer, supposedly by wrapping around the tubular aggregate. For the latter, growth of a second dye double layer of the aggregates seems to grow. The different morphologies of the coatings are rather due to the details of the chemical structure of the polycations than due to physical parameters. The possible adsorption of polyelectrolytes at these amphiphilic tubular structures, stabilized by non-covalent bonds only, is far from obvious but demonstrates an applicable route to the build-up of more complex nanostructures by means of a self-assembly process.

# Introduction

The fabrication of functional units on mesoscopic length scales (nanometers to micrometers) by a self-assembling process is a fascinating but challenging research topic [1,2]. It can be considered as a biomimetic approach following design rules of living biological matter. Therefore aqueous environment is a prerequisite to utilize electrostatic and hydrophobic forces for the combination of a variety of materials on the base of elementary physical laws. For tailor made construction of functional systems, it is highly desirable to be able to combine different materials with different functionality in a step by step method.

One practical and well established method to achieve this is layer-by-layer adsorption of oppositely charged macromolecules on colloidal systems. [3-7] Although this method was initially introduced for adsorption of polyelectrolytes (PEs) on planar surfaces [8], it was successfully applied to spherical [9,4] and cylindrical [10] colloidal objects as well. Colloidal objects of various materials such as polystyrene latex [4], melamine formaldehyde [3], silica [6], or gold [11,12] and various sizes from microns down to a few nanometers were employed as substrates. The key feature for successful adsorption of one or more layers of polyelectrolytes is charge overcompensation leading to inversion of surface charge. [8,13,5] The charge reversal is also necessary to maintain the stability of the colloidal suspension. Experimentally, the charge reversal was verified by electrophoretic measurement of the zeta-potential [4,14].

The amount of adsorbed polyelectrolyte depends on several parameters in addition to surface charge density of the colloidal species, such as ionic strength of the solutions, pH of PE solution, adsorption time, temperature, PE concentration, charge density of PEs, PE rigidity (persistence length), PE's molecular weight, and curvature of the colloidal particles. One of the most critical parameters is salt concentration or ionic strength of the PE solution, because it significantly influences the conformation of Pes in solution [15]. At low salt concentration the PE molecules adopt a stretched conformation due to electrostatic chain stiffening [15]. In that case charge overcompensation of a substrate is difficult to be achieved, because the PE chains are assumed to lie almost flat on the surface.[16] At high salt concentration the PE is in a more coiled state with a reduced persistence length, and hence reduced radius of gyration. In that case the adsorption can be considered as a process where the coil size remains almost unchanged during adsorption and the PE hits the surface as a clot ('hit-and-stick' model [5]). Coverage of the surface is then obtained by filling with these clots and the adsorbed amount of a complete layer is expected to scale with ionic strength [17,18]. Overcompensation of surface charge is then easily obtained and the thickness of the adsorbed layer is of the order of the dimension of the coil size in solution. Besides the attractive electrostatic forces also repulsive forces between segments of the adsorbed PEs [15] have to be considered as well as entropic effects [19,20] due to ion release [5] and due to hydrophobic effects caused by local changes of entropy within the water.

For the adsorption of polyelectrolytes on colloids, there are further limitations to avoid precipitation, which can be derived from simple geometrical considerations. During the coating process there is competition between the formation of a monolayer on the colloid and coagulation of partly covered, hence less charged, particles. This implies that the adsorption process has to be fast compared to the mean time where two particle meet. These time scales can be controlled by the mean distance between colloids and between colloid and polyion, i.e. by the respective concentration [21]. The molar polymer concentration should be large compared to the colloid concentration and a larger amount of polymer than needed to completely cover the colloidal surface should be present in order to make adsorption of PE faster than agglomeration of colloids or partially covered colloids.

On the other hand the mean distance between the colloidal particles has to be larger than the typical extension of the PE chains to avoid bridging of two or more colloidal particles by one chain, which would facilitate flocculation. Therefore, the total concentration of colloids has to be rather low.

So far, in most of experimental work and all theoretical investigations known to us the colloidal objects coated by PEs were considered to be rigid and with fixed (or pH-controlled) surface charge density. Only few experiments of polyelectrolyte adsorption have been reported for soft materials that are held together by hydrophobic and other non-covalent interactions (association colloids), such as micelles [22,23] or cells[7]. In this case the adsorption of polyelectrolytes can easily modify the molecular structure. E.g., it was shown for lipid structures that the PE adsorption causes reorientation and flipping of the lipids within the double layer membrane [24]. A more stable adsorption without disturbance of membrane integrity was observed for slightly hydophobized polycations, or for liposome membranes in a liquid crystalline state, or for large liposomes with diameters of several hundreds of nanometers.[24] Therefore, it is challenging but not impossible to build-up complex nanostructures by hierarchic self-assembly of differently charged macromolecules on liposome-like structures. For controlled assembly of polyelectrolytes on such liposome-like structures a system would be desirable that allows easy monitoring of the integrity of the molecular packing within the structure before and after the adsorption process.

In this work we use tubular J-aggregates of the amphiphilic cyanine dye 3,3'-bis(2-sulfopropyl)-5,5',6,6'-tetrachloro-1,1'-dioctylbenzimida¬carbo¬cyanine (C8S3) as a liposome-like association colloid. The dioctyl-substituted dye molecules adopt a molecular packing structure similar to a lipid bilayer, which then bends and closes to form a nanotube [25-27]. The inner and outer diameter of the tube are 6.5±0.5 nm and 13±1 nm, respectively. The optical spectra exhibit the typical features of J-aggregates, [28,29] namely, new transitions that are red shifted with respect to the monomer absorption and significantly narrowed due to collective excitations [25]. The spectrum of C8S3 aggregates is dominated by two pronounced peaks at 589 nm and 599 nm, two smaller peaks at 557 nm and 570 nm, and a shoulder at 582 nm (see Figure 1), all red shifted with respect to the monomeric absorption at 540 nm. This particular spectral pattern is due to the bilayered tubular structure of the aggregates and explained and described in detail in the literature [30-32]. The two strong peaks are assigned to excitons located within the inner wall (599 nm) and the outer wall (589 nm), respectively [25-27,33]. Small structural changes of the molecular order within the aggregate cause either wavelength-shifts of the peak maxima or changes in the relative intensity of the peaks. From the , which also can be assigned to the outer or inner region of the tube wall. Therefore absorption as well as photoluminescence spectra serve as a sensitive, finger-print like, indicator for the structural integrity of the tubular aggregates.

The adsorption of three different polycations at the surface of the dye aggregates is studied: Polyethylenimine (PEI), poly(allylamine hydrochloride) (PAH), and poly(diallyl-dimethylammonium chloride) (PDADMAC). Two of the polycations (PAH and PDADMAC) are linear PEs with similar line charge density in fully charged state, but different persistence length and different amounts of hydrophobic elements [34,35], while the third one (PEI) is a linear PE with many short branches that lead, in the fully charged state, to an increased line charge density[36]. Because of the stability requirements for the aggregates the variability of solvent parameters is rather limited. On the one hand, the concentration of dye must be higher than about $10^{-4}$ M to obtain a sufficient amount of stable aggregates to be observable by TEM, and no additional salt should be used to avoid bundling of the tubes (see supplementary information). That means we are in the limits of high colloid and low salt concentrations. On the other hand, the polyelectrolyte concentration has to be rather low to avoid bridging between the tubes and hence precipitation. Thus we will not be able to reach the required excess of polyelectrolytes. We investigate the morphology of the adsorbed polycation/dye-aggregate complexes by cryo-TEM and follow the integrity of the aggregates by UV/Vis spectroscopy. The most regular wrapping is found for the linear polyelectrolyte PDADMAC. For this case a thin homogeneous covering is obtainable that can be made visible by staining with negatively charged inorganic nanoparticles. In case of PEI the appearance of clots is dominant indicating a hit-and-stick adsorption of globules, and in case of PAH an irregular structure is obtained with deformation or even destruction of the outer dye layer. The adsorption of polyelectrolytes was used by us in recent work to add semiconducting nano particles that serve as donors or acceptors in Förster energy transfer [37] which shows that electrostatic adsorption on these aggregates is a useful tool for the build-up of higher organized structures. But still this is the first more systematic study of the adsorption of linear polyelectrolytes on a lipid-like quasi one-dimensional tubular structure according to our knowledge.

## Materials and Methods

**Materials.** Cyanine dye 3,3'-bis(2-sulfopropyl)-5,5',6,6'-tetrachloro-1,1'-dioctylbenzimida-carbocyanine (C8S3, MW = 902.8 g/mol, see Scheme 1) was purchased from FEW Chemicals (Dye S 0440, FEW Chemicals, Wolfen, Germany) and used as received. (Note: according to recent findings this product contains significant amount of iodine as counter ions [38]) Poly(allylamine hydrochloride) (**PAH,** average $M_w$ = 17 000 g/mol, 20 wt% in $H_2O$), polyethylenimine (**PEI,** average $M_w$ = 750 000 g/mol, 50 wt % in $H_2O$), and poly(diallyldimethylammonium chloride) (**PDADMAC,** 20 wt% in $H_2O$) have been purchased from ALDRICH and were used without further purification. PDADMAC with two different molecular weights was used: average $M_w$ = 100 000 – 200 000 g/mol (PDAD100) and average $M_w$ = 400 000 – 500 000 g/mol (PDAD400). For all aqueous solutions ultra-pure deionized water was used (Millipak Express 20, 0.22 µm filter, σ= 18 MΩcm).

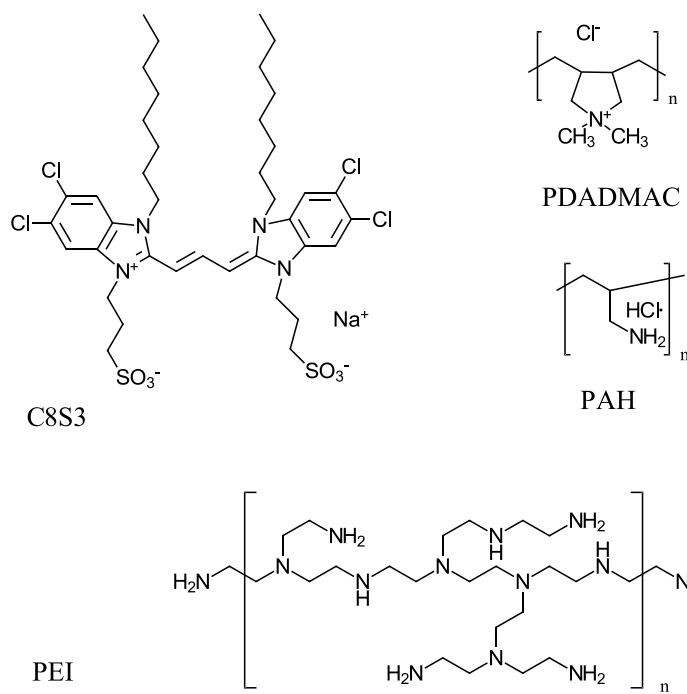

**Scheme 1**. Chemical structure of amphiphilic cyanine dye C8S3 (3,3'-bis(2-sulfopropyl)-5,5',6,6'-tetrachloro-1,1'-dioctylbenzimida-carbo-cyanine) and the polycations PEI, PAH, and PDADMAC.

**Preparation.** For the preparation of C8S3-aggregates the so called *alcoholic route* was followed [26]: C8S3 was dissolved in pure methanol under vigorous stirring to obtain a 2.92 mM monomeric stock solution. In order to prepare C8S3 J-aggregates 130 µl of the C8S3 stock solution were added to 500 µl of ultrapure H$_2$O. The solution was stored in the dark for 24 h before adding an additional 500 µl of H$_2$O resulting in a final dye concentration of [C8S3] = 3.36 · 10$^{-4}$ M and a total relative amount of 11.5% of methanol. It was found that aggregated solutions prepared by this procedure are stable over periods of several days in the sense that no qualitative change of the absorption spectra is observed [39] (see supplementary information). All glass vials were stored in distilled water for more than 48 h before used for dye or dye aggregate solutions. The solutions of J-aggregates were used for experiments within three days of preparation and permanently stored in the dark. For most of the wrapping experiments polyelectrolyte solution (PE) of various concentration (throughout the text polyelectrolyte concentrations are given by the molar concentration of monomers) was added to aggregated dye solution in a 1:1 volume ratio. Sometimes volume ratios of 10:1 (dye to PE solution) were used; no effect on the results was observed that originated from different volume ratios of the solutions. The mixtures were then shaken by hand for 10 s and stored in the dark.

**Characterization.** Absorption spectra were taken by a SHIMADZU UV2101 spectrophotometer. Samples were measured in 1 mm quartz-cuvettes, if not stated otherwise. Absorption measurements were corrected for solvent background. Zeta potential measurements were carried out on a Zetasizer Nano ZS analyzer with integrated 4 mW He-Ne laser, $\lambda = 632.8$ nm (Malvern Instruments Ltd, U.K.). To elucidate the surface charge of aggregates and of coated aggregates, measurements were carried out in phosphate buffer. 100 ml buffer was prepared using 4.27 ml of 0.02 M $KH_3PO_4$ and 30.49 ml of 0.01 M $Na_2HPO_4$. Additional 65.24 ml of pure Milli-Q water (Millipak Express 20, 0.22 µm Filter) have been added to fill the volume, resulting in a pH-value of 7.6. All measurements were carried out at 25 °C using folded capillary cells (DTS 1060) in triplicate.

For cryo-TEM investigations the samples were prepared on Athene mesh 200 copper grids covered by Quantifoil® film (1/4) (Quantifoil, Germany), hydrophilized by glow-discharge in a BalTec MED020 for 60 s. A small droplet of the sample was dispersed on the grid and blotted immediately before shock - freezing in liquid ethane [39,40]. Vitrification of the ultrathin aqueous layers spanning the holes of the Quantifoil film was achieved by the high cooling rates of the liquid ethane and the fast immersion using a standard plunging device. The vitrified samples were stored under liquid nitrogen until transferred into a Philips CM12 TEM with cryo-holder (Model 626, GATAN). TEM-images were obtained in bright-field low dose mode using a defocus of 1.2 to 1.4 µm.

# Results

## Boundary conditions for sample preparation

The adsorption of the polycations to the tubular J-aggregates of C8S3 is studied by absorption spectroscopy and cryo-transmission electron microscopy (cryo-TEM). In general, the samples are prepared by adding an aqueous polycation solution with concentration (with respect to the monomers) much less than the C8S3 concentration ($3.3 \cdot 10^{-4}$ M, in an 100:13 water:MeOH mixture). Either equal volumes of polycation solution are added or smaller volumes with respectively higher concentration are added. No effect was observed neither with respect to the spectra nor with respect to the cryo-TEM images that would favor the one method over the other. Since the tubular aggregates are self-assembled structures there are certain constraints for experimental parameters regarding the solvent (details are outlined in the SI):

First, there is a minimum concentration for the formation of aggregates defined by the critical micellar concentration (at room temperature) which is $(4.2 \pm 0.6) \cdot 10^{-5}$ M. Below this concentration no aggregates are found. For practical reasons a concentration of $3.3 \cdot 10^{-4}$ M was chosen, because at lower concentrations only very few aggregates could be found by cryo-TEM.

Second, there is a limited time window over which the isolated aggregates are stable. The process of aggregate formation takes approximately 24 h and after two days they start to agglomerate into bundles. Although these time effects can be easily taken into account, they show that the aggregates are not equilibrium structures but represent a rather metastable state [41]. The bundling of the aggregates is dramatically enhanced by addition of salts in concentrations higher than the dye concentration (see SI). This effect is obviously due to Debye screening of the electrostatic repulsion of the negatively charged tubular aggregates similar to the case of spherical colloids described by the DLVO theory [42].

Third, the dyes are affected by the pH of the solution. In the acidic range with pH < 4 the dyes are colorless, probably due to protonation of the polymethine group of the chromophore.

In the basic regime with pH > 13 the dyes precipitate. Hence there is a pH window between 4 and 13 available for experiments.

All experiments described in the following were performed in a pH range between 7 and 9 and without any additional salt. Samples were freshly prepared by using solutions not older than two days if not stated otherwise. We will now present the results for every polycation individually and then discuss the results in total.

**Polyethylenimine (PEI)**

The influence of the polyelectrolyte on the aggregate structure was primarily checked with the help of the absorption spectra. **Figure 1** presents changes of absorption spectra upon addition of PEI solutions at various concentrations. No additives were used to adjust the pH value. For the given $pK_a$ of PEI of 8.2 [36] a pH value of 9 is calculated for the PEI/dye solutions. As shown in **Figure 1**, the addition of PEI in concentrations below $10^{-5}$ M causes only slight reduction in absorbance and very small red shifts of less than 1 nm of the absorption maxima of the strongest transitions. For PEI concentrations above $10^{-5}$ M the absorbance is significantly reduced and the peaks become broadened. This broadening is typical for a high degree of disorder within the J-aggregate system [43]. The decline of absorbance is accompanied by precipitation of large flakes that are visible by naked eye. For a concentration of $10^{-4}$ M the J-bands are still weakly observable, although the intensity loss is drastic. For a concentration of $10^{-3}$ M the material completely precipitates and no absorbance is observable anymore. Therefore, the coating properties of PEI were investigated at PEI concentrations in the range of $10^{-5}$ to $10^{-6}$ M, which corresponds to a ratio of PEI monomers to dye monomers of 1:30 to 1:300. The amount of PEI is far too little to allow for a complete coverage of all of the aggregates present in solution and does not fulfil the proposed requirement of excess of polyelectrolyte to achieve charge reversal upon adsorption at an oppositely charged surface [15,44].

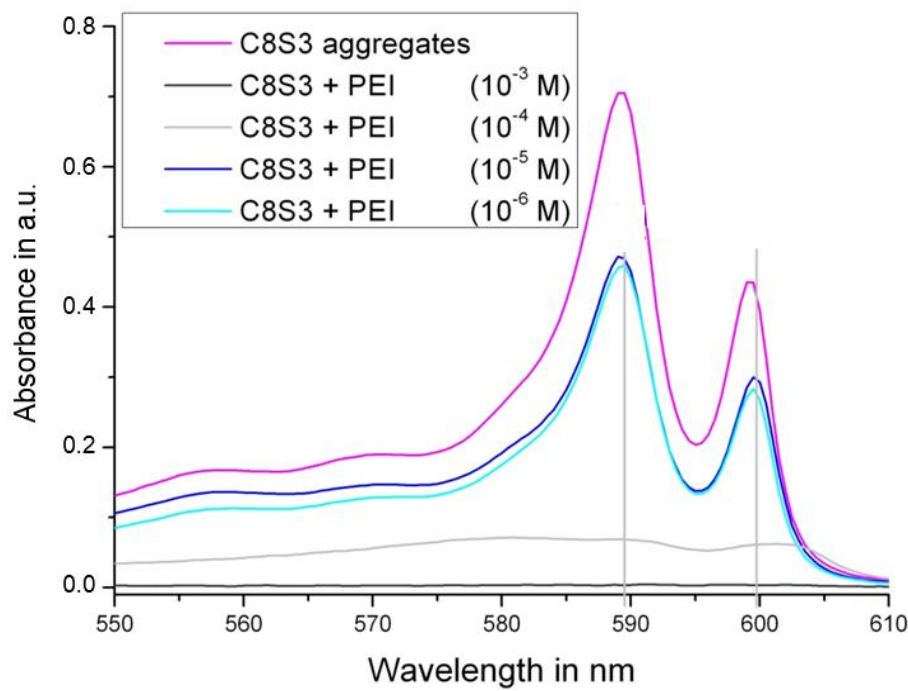

**Figure 1**: Absorption spectra of C8S3 aggregates after addition of PEI in different concentrations. The C8S3 concentration was $1.7 \cdot 10^{-4}$ M. The vertical grey lines are guiding the eyes and mark the peak positions. (Shown here is the part of the spectrum that is relevant for the interpretation. At lower wavelengths the spectrum decays monotonously to zero at about 330 nm. More details are found in the SI).

**Figure 2** displays cryo-TEM images of the aggregates with PEI added in a concentration of $10^{-5}$ M. One can see that most of the aggregates are unaffected and appear as homogeneous and uniform tubular structures with an outer diameter of $6.5 \pm 0.2$ nm, as they were observed in pure solutions of C8S3 at same conditions [45,25]. However, some of the aggregates are irregularly decorated with clots which are assigned to PEI (see **Figure 2**a).). Some parts of the aggregates appear to be rather homogeneously covered (see **Figure 2**b) with a layer of about 2 to 4 nm. The average total thickness of the adsorbed material is estimated to be in the range of 4 to 6 nm and the fraction of aggregates that is coated by PEI as estimated from several cryo-TEM images is less than 10%. The appearance of PEI here is very similar to what was observed by Liu et al. for PEI adsorbed at carbon nanotubes [46]. They convey the impression that most of the PEI is adsorbed from its coiled conformation in solution and simply sticks to the aggregates surface ('hit-and-stick adsorption' [5]).

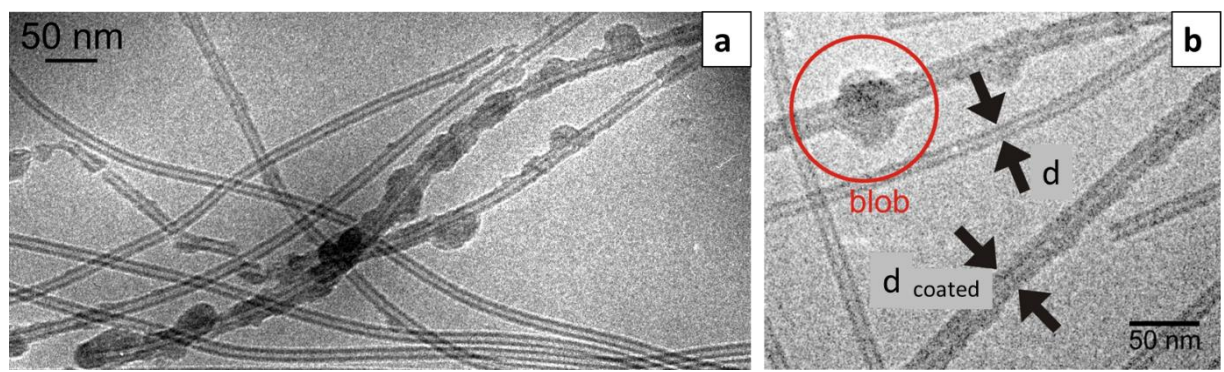

**Figure 2.** Cryo-TEM images of a vitrified solution of C8S3-aggregates after addition of $10^{-5}$ M PEI solution. (a) Overview showing predominantly uncovered tubular aggregates; (b) Image showing a clot ("blob") of PEI adsorbed to an aggregate and tubular aggregates that are homogeneously coated. The thickness of the uncoated (d = 13±0.5 nm) and the coated aggregate (dcoated = 18±1 nm) are marked with arrows.

**Poly(allylamine-hydrochloride) (PAH)**

Already low concentrations of poly(allylamine-hydrochloride) (PAH) added to solutions of C8S3 aggregates result in strong qualitative changes of the absorption spectra as shown in **Figure 3**.

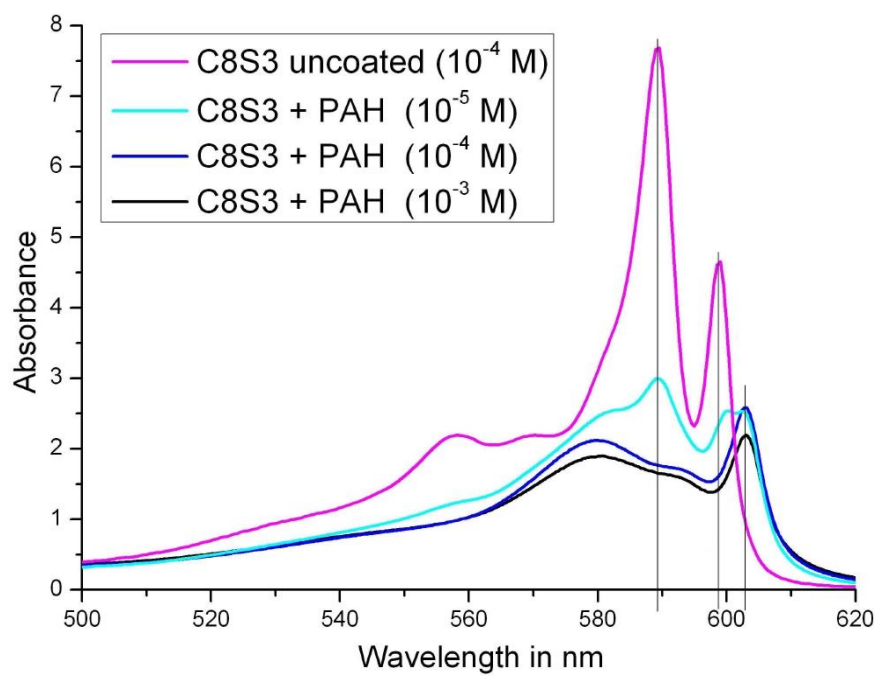

**Figure 3:** Absorption spectra of C8S3-aggregates after addition of PAH at different concentrations. The vertical grey lines are guidelines for the eyes to mark the peak positions.

The initially very prominent absorption bands at 589 nm and 599 nm of the outer and inner wall, respectively, vanish completely at higher concentrations of PAH (above $10^{-4}$M) and new bands centred at 578 nm and 603 nm dominate the spectrum. Additionally, the small bands at 560 nm and 570 nm vanish. The band at 580 nm may be the remaining of the shoulder at this position of the original spectrum. These significant changes in the absorption spectra are due to changes of the molecular organisation within the tubes and are not understood in detail. However, they are very similar to the spectral changes that are observed upon bundling of the aggregates [25,41] (see also SI). Thereby the shift of the band at 599 nm to 603 nm is particularly indicative of a slight structural change of the inner tube wall, while the vanishing of the peak at 589 nm indicates destruction of the outer wall. It was shown by Eisele et al. [41] that the bundling essentially removes the outer layer of the tubes, resulting in a structure of packed tubes that only consist of the slightly modified inner tube, sheathed by an outer layer that encloses the entire bundle. It is therefore reasonable to assume that the PAH coverage destroys primarily the outer layer of the aggregates. In addition to the shift of the absorption bands, the absorption strength is significantly reduced. The resulting absorbance becomes almost independent of the PAH concentration at higher concentrations. This can be interpreted to mean that at concentrations above $10^{-4}$ M PAH (roughly 1:1 concentration ratio between PAH monomers and dye molecules) all aggregates are covered.

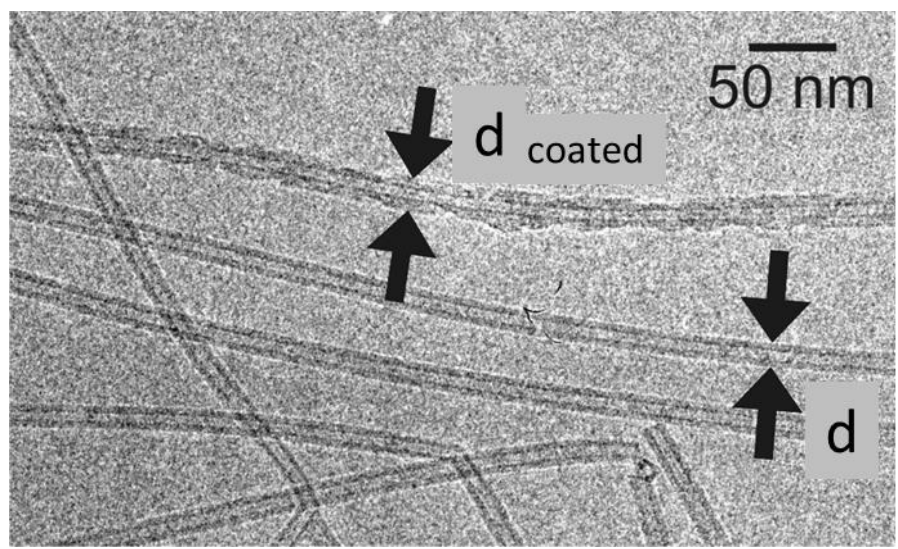

**Figure 4.** Cryo-TEM image of a vitrified solution of PAH-wrapped C8S3-aggregates; PE-concentration $10^{-5}$ M. The thickness of the uncoated (d = 13±0.5 nm) and the coated aggregate (dcoated = 18±1 nm) are marked with arrows.

For cryo-TEM imaging samples were prepared from $10^{-5}$ M PAH solution, which corresponds to a dye:PAH monomers ratio of 30:1. The cryo-TEM image of **Figure 4** shows uncoated aggregates in coexistence with an aggregate coated with PAH. A survey of a large number of TEM images (see SI for more images) reveals that aggregates are either coated in total or not at all, as is apparent from **Figure 4**. For samples with $10^{-5}$ M PAH solution added less than 20% of the aggregates show coatings. The coating itself appears quite homogeneous along the aggregates but rippled or undulating. In some places, the images create the impression of a coating helically wound around the aggregate. The mean thickness of the coating is 2.5 nm ± 0.7 nm. On aggregates that are coated, the inner wall of the tube can be identified, but the outer wall is no longer visible, which is suggested to be an indication for the disintegration of the outer wall. Sometimes aggregates were found where the tube structure appeared totally destroyed. Those structures look like a creased tape with a width similar to the original tube diameter. One example of such a structure is shown in the SI (Figure S1).

The amount of PAH relative to C8S3 could be increased to a ratio of approx. 3:1 by adding PAH up to a concentration of $10^{-3}$ M. For that case the TEM-images reveal a high ratio of coated aggregates of more than 80%. However, the tubular structure for these coated assemblies is no longer observed, instead, the tape-like assemblies as found for lower PAH concentrations prevail (see Figure S2).

**Poly(diallyldimethylammonium-chloride) (PDADMAC).**

The addition of PDADMAC in $10^{-6}$ to $10^{-5}$ M concentrations (which corresponds to a molar ratio PE:dye of 1:3300 to 1:330) to a freshly prepared C8S3-aggregate solution neither generated any change of the absorption spectra nor did it cause turbidity of the solution or precipitation. **Figure 5** shows absorption spectra of a pure C8S3 aggregate solution and with PDADMAC at concentration of $10^{-6}$ M and $10^{-5}$ M. The addition of higher concentrated PDADMAC solutions however leads to precipitation of reddish flakes over a certain period of time. For a concentration of $10^{-4}$ M it took a few minutes before visible precipitation started, but for a $10^{-3}$ M PDADMAC solution this was observed immediately after addition of the polyelectrolyte to the dye solution. This threshold for precipitation was similar for both molecular weights of PDADMAC (PDA100 and PDA400). The absorption spectra spectra did not change qualitatively, which means that the molecular arrangement within the dye aggregates is not altered by the adsorption of the PDADMAC.

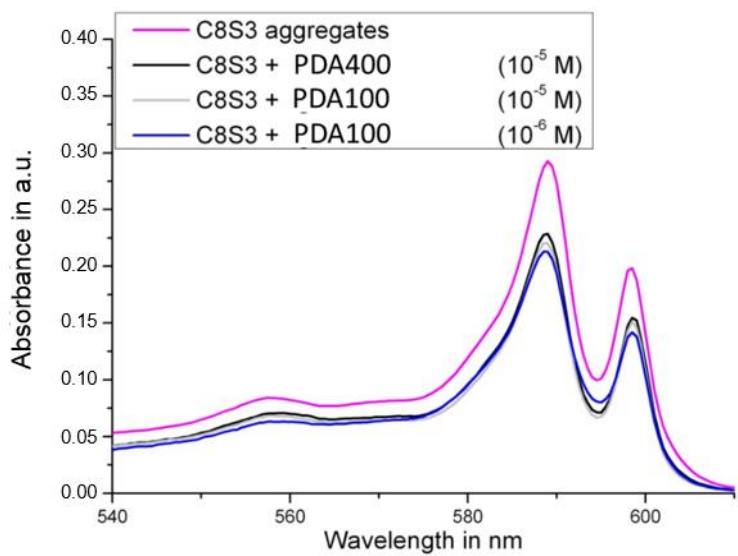

**Figure 5**: Absorption-spectra of aggregates of $3.3·10^{-4}$ M C8S3 solution after addition of PDADMAC in different concentrations. The spectra are corrected for concentration differences due to mixing with PDADMAC solutions.

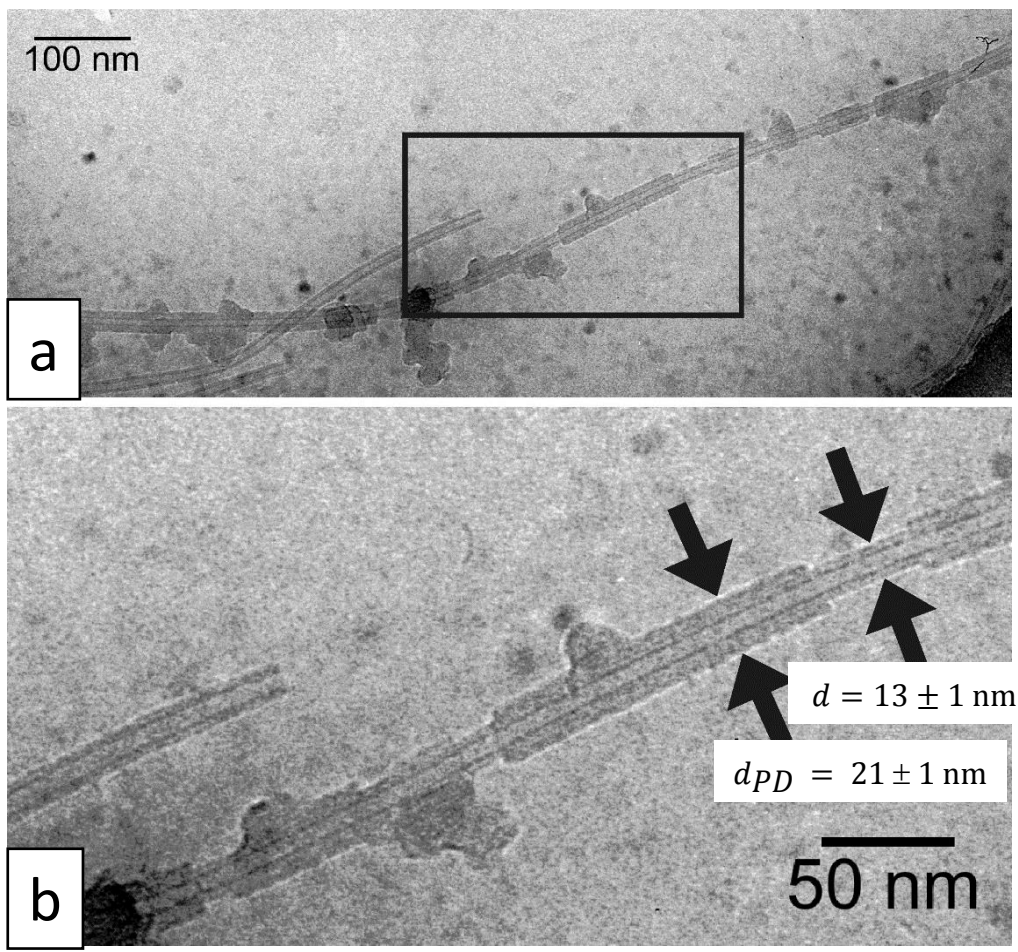

**Figure 6.** Cryo-TEM images of $3.3 \cdot 10^{-4}$ M C8S3 solution after addition of $10^{-4}$ M of high molecular weight PDADMAC (PDA400) solution. In (b) a magnified view of (a) as marked by a frame is shown. See text for more information.

Cryo-TEM images were taken from samples of C8S3 prepared by mixing with $10^{-4}$ M solution of PDADMAC. Typical images are shown in **Figure 6** for the high molecular weight PDADMAC (PDA400). Very similar images were obtained for the low molecular weight PDA100, see SI. As in the case of PEI and PAH one finds bare aggregates with a typical outer diameter of $d$ = 13 nm ± 1 nm. Other aggregates are decorated by clot-like structures and some sections of aggregates have an increased width of the tubular wall resulting in an outer diameter of $d_{PD}$ = 21 nm ± 1 nm, see **Figure 6**. The diameter of the inner part of the tubular aggregates always remains unaffected. Very often the segments with increased wall thickness are accompanied by blobs.

As in the case of PEI and PAH we ascribe the blob like structures to clots of pre-assembled PDADMAC. The typical size of these clots is in the range of 10 to 50 nm which fits well to measured radii of gyration (22 nm to 40 nm) of the globules in solution for comparable PDADMAC under comparable conditions [35]. The homogeneous thickening of the wall at the corresponding segments may be explained by the formation of an additional tube wall. The regular tube wall consists of a dye double layer, comparable to a lipid double layer with both sides charged negatively. Assuming that the outer layer of the tube is covered by cationic PDADMAC causing charge reversal, one may presume the growth of another double-layer of dyes on top of the regular tube wall. One indication for such an interpretation is the increase in thickness and the homogeneity of these segments: the thickness of one dye double layer of a bare aggregate (i.e. the tube wall) is $3.2\pm0.5$ nm, while the increased wall thickness amounts to $7.3\pm0.6$ nm, which is just a bit more than twice the thickness. But further evidence for this doubling of the tube wall is obtained from staining experiments. In a previous publication [37] we described the adsorption of negatively charged CdTe nanoparticles at aggregates that were coated with PDADMAC. The CdTe nanoparticles can easily be detected by cryo-TEM and therefore serve as a staining agent for the electron microscope. Since they are attracted only by positive charges they stick only to those parts of the aggregates which are covered by PDADMAC. The particles are found attached to the clots visible in **Figure 6** and on those parts of the aggregates that are not appreciably thickened, but they are not found on the segments with almost double wall thickness. Thus, one may conclude that these segments are negatively charged and not covered by PDADMAC, and that at least part of the remaining aggregates are be wrapped with a very thin layer of PDADMAC. Due to the small number of aggregates observed by cryo-TEM, no statistically significant statement about the proportion of coated or thickened aggregates is possible. However we want to mention that only a small fraction of the total length of the aggregates show such doubling of the tube wall and therefore any spectral changes caused by this new dye layers are not visible in the spectra.

## Discussion

Although the appearance of the coatings and their effects on the optical spectra of the aggregates is very different for the different PEs, there are two interesting observations that are in common for all of them: First, not all aggregates are covered with polyelectrolyte, but second, individual aggregates were either at most totally covered with adsorbed PE or totally uncovered.

The incomplete overall coverage of the aggregates is due to the insufficient amount of polyelectrolyte material added to the aggregate solution, since in all experiments the concentration of monomer units of the PE is lower compared to the dye concentration by a factor 10 to 100. Even for a 1:1 adsorption ratio of monomer to dye one would expect to observe adsorbed PE on only about 1% to 10% of the aggregates. This can be estimated more precisely with the help of a continuum model with tubes of inner radius of $r_i = 3.2$ nm and outer radius of $r_o = 6.5$ nm that are surrounded by a PE shell with radius $r_{PE}$. From the TEM images, the fraction of length of aggregates that is coated with PE $f_{L,exp}$ is taken as the ratio of overall length of coated aggregates $L_{PE}$ compared to the length of all aggregates $L_A$

$$f_{L,exp} = \frac{L_{PE}}{L_A}.$$

This length ratio is converted into a volume ratio using the average thicknesses of the coating $r_{PE} - r_0$ by

$$f_{V,\exp} = \frac{V_{PE}}{V_A} = \frac{L_{PE} \cdot (r_{PE}^2 - r_o^2)}{L_A \cdot (r_o^2 - r_i^2)} = f_{L,exp} \cdot \frac{(r_{PE}^2 - r_o^2)}{(r_o^2 - r_i^2)}$$

The same volume ratio can also be calculated from the mass concentration ratio of C8S3 and the respective PE taking into account their respective molecular volume $v$ (see Table 1):

$$f_{V,\text{calc}} = \frac{[PE] \cdot v_{PE}}{[C8S3] \cdot v_{C8S3}}.$$

Given that, one can calculate the length ratio $f_{L,calc}$ of covered aggregates for the respective average thickness of coatings which can be compared to the experimental value. The results are summarized in Table 1 with molecular volumes taken from computer models [27] or, in case of C8S3, from x-ray data analysis [47]. The calculated fraction of covered aggregates is always smaller than the fraction estimated from the cryo-TEM images, which however suffers from poor statistics due to small sample size for TEM images. For PDADMAC the fraction of covered aggregates cannot be taken from the images, since the thin coating is not visible by cryo-TEM. The calculation was done for the assumption of a coating thickness of 1 to 2 nm; the resulting high degree of coverage for this relatively high concentration of PDADMAC is in accordance with observations of stained aggregates [37].

Table 1: Data for comparison of coating volumes.

| Material | $v$ / nm$^3$ | [c] / mol/l | $r_{PE}$ | $f_{L,exp}$ | $f_{L,\,calc}$ |
|---|---|---|---|---|---|
| C8S3 | 1.1 | 3.4·10$^{-4}$ | | | |
| PEI | 0.09 | 0.7·10$^{-5}$ | 8.5 … 9.5 | < 10% | 2% … 1% |
| PAH | 0.13 | 0.75·10$^{-5}$ | 97.5 … 8.5 | < 20% | 6% … 3% |
| PDADMAC | 0.26 | 6.4·10$^{-4}$ | 7.5 … 8.5 | Unknown | 95% … 44% |

The coating-ratio could be increased to some extent by adding PE-solutions with higher concentrations. Attempts with PAH were successful up to concentrations of 10$^{-3}$ M, providing coating-ratios of approximately 80%, however at the cost of the integrity of the aggregates. These high coating ratios without precipitation were only obtained for PAH that was purified by chromatographic methods. The zeta-potential measurements in that case revealed more positive surface-potentials.

The coexistence of modified and unmodified aggregates was observed also for coverage of the aggregates with inorganic materials [48,49] even for the growth of AgI crystals within the aggregates [50]. Therefore, this seems to be a typical feature of the aggregate system and may be explained by the dynamics of the aggregates due to the amphiphilic nature of the molecules. It is well-known that self-assembled aggregates of amphiphilic systems such as micelles, lipid bilayers, or vesicles, are in a state where molecules can exchange between solution and aggregates [51,42]. The growth of the aggregates over a timescale of minutes after dilution of the alcoholic stock solution (see SI) shows that there is dynamic exchange with monomers. In equilibrium, the monomers are present in critical micelle concentration (cmc) which is about $10^{-5}$ M in our case. Hence, we assume that there is continuous exchange between monomers in solution by dissolution and growth of aggregates, which is slow enough to allow adsorption of PE on existing aggregates, but fast enough to destroy and re-build over time remaining parts of aggregates that are not covered by PE. Now if the coverage hinders dissolution of aggregates, one ends up with an equilibrium between covered and non-covered aggregates instead of partially covered aggregates. However, we have to admit that so far there are no systematic experimental studies about this dynamic exchange of molecules.

The most interesting observation is the different geometry of the coating for the three different polycations. A smooth and thin coating is characteristic for PDADMAC, while a lumpy coating is characteristic for PEI. A more regular but rough coating is found for PAH. The adsorption behavior of single chains of polyelectrolytes on oppositely charged surfaces is treated intensively by theory [15,52-55] and the relevant physical parameters of the polyelectrolytes in that case are line charge density and mechanical persistence length. The conditions for adsorption and the manner of wrapping of the PE chain depend on the geometry of the colloid, the surface charge density, and the Debye screening length given by ionic strength of the solvent. The physical parameters of the polycations used here are quite similar (at least within the same order of magnitude, see Table S1) even for different states of dissociation of the weak polyelectrolyte PAH. Therefore, from the theoretical point of view, one would expect a very similar behavior for the three polyelectrolytes.

The differences of the polycation behaviour merely originates from their molecular structure than by the physical parameters. The polycation polyethylenimine (PEI) is a linear polyelectrolyte with short branches on almost every second monomer with primary, secondary, and ternary amine groups (see **Scheme 1**). The structure of commercially available PEI was characterized in detail by van Harpe et al. [36] and they found for PEI with the same molecular weight used here an almost equimolar amount of each of the amines (35:35:31 for primary:secondary:ternary amine). Therefore, nearly every second nitrogen forms a branch. For the $pK_a$ value of PEI of 8.2 on average, the pH of the aqueous $10^{-5}$ M PEI solution is calculated to be 9. At this pH the degree of protonation of PEI is on the order of 20%. [56-58] It would be fully protonated at low pH (<4), which is not accessible for the aggregates because it would lead to protonation and hence decoloring of the dyes (see SI). Therefore the PEI can be assumed to be present in a rather coiled state, and its radius of gyration can be estimated from literature values, which was given for a PEI with molecular weight of 581,000 g/mol and in 1 M NaCl solution to be 22 nm [59]. Therefore, in the salt free case here it will be more extended, but presumably not completely stretched due to the low charge density.

PEI is known to be a very strong adhesive on a variety of surfaces and therefore is often used as the first layer in the preparation of polyelectrolyte layer-by-layer assemblies due to the strong adhesion to different substrate materials [8]. Typical effective thickness of a single PEI layer adsorbed on glass surface from aqueous solution at neutral pH was reported to be in the range 3.5 to 6 nm [60], but with a roughness of the same order of magnitude. Scanning force microscopy studies of adsorbed monolayers of PEI reveal a homogeneous coverage with globules smaller than 60 nm [61]. These data are in accordance with the behavior of PEI we observed here, namely, the adsorption of PEI chains that in solution exist in a coiled state, forming a clot-like structure at the surface. The chains do not unfold and wrap around the cylindrical tube but adsorb in a "hit-and-stick" manner.

The PAH and the PDADMAC are linear chain polyelectrolytes but with different amine groups. PAH is a weak PE that is not fully protonated at neutral pH [17], while PDADMAC with its quaternary amines is a strong polyelectrolyte that is positively charged in neutral to slight basic pH range. Other differences are the flexibility of the chains and the amount of hydrophobic parts of the chemical structure. The difference in flexibility, i.e. persistence length, may be seen from the radii of gyration obtained from light scattering for chains with comparable molecular weight (see Table 1, SI). They are larger by a factor of almost five for PDADMAC compared to PAH. Consequently, PAH may be considered as the more flexible chain. The high flexibility of PAH facilitates complex formation between monomers of PAH and dye molecules and therefore may be responsible for the destruction of the dye tube upon adsorption.

To explain the adsorption in a thin and homogeneous layer of PDADMAC, which is very different from the behavior of PAH, we can only speculate. One reason could be hydrophobic effects, as they are observed by temperature dependent multilayer adsorption. [62] One quantity used to describe "hydrophobicity" of polyelectrolyte chains was the number of carbon atoms per monomer unit, which is 8 for PDADMAC and 3 for PAH. Additionally, in PDADMAC the carbons are within the five-membered ring, which increases the hydrophobicity. Investigations on the adsorption of polycations on anionic liposomes presented in [24] revealed that the binding of the PE could be enhanced by hydrophobization of the PE with additional chemical groups. This effect is not fully understood yet.

As is seen by the spectra, the molecular order of the aggregates is, at least for PDADMAC, almost not disturbed by the adsorption of the PE. This indicates that the binding of the PE to the aggregate surface is less strong than the intermolecular binding energy between the dyes within the aggregate. Again, from adsorption studies on liposomes it was found [24] that adsorption of PE is enhanced if the liposome membrane is in a liquid crystalline state. In our system the dyes are even more ordered [27] compared to the hexatic liquid crystalline states of lipid membranes [42] which may additionally contribute to the strong adsorption of PDADMAC.

## Conclusions

In summary, we successfully coated tubular J-aggregates of anionic amphiphilic cyanine dyes with three different polycations. The different chemical nature of the polyelectrolytes results in different appearances of the coatings. The branched polycation PEI attaches to the tubular aggregate by hit-and-stick adsorption of the coiled state in solution, the flexible and weak polycation PAH forms more closed layers on the aggregate surface but destroys the integrity of the molecular aggregate, while the more hydrophobic and strong polycation PDADMAC forms a thin and homogeneous layer. There is evidence that this polycation wraps around the aggregate in a controlled manner.

We conclude that adsorption of polyelectrolytes at these amphiphilic tubular structures, held together only by non-covalent interactions, is a feasible way to build more complex nanostructures. Such hierarchic assembly makes it possible to combine different materials, each with different functionality, into a new unit, in which the functions of the individual parts have to fit together precisely to fulfill a completely new task. As a simple example of a first successful step in this direction one may consider the energy transfer units described in [37], where the C8S3 aggregates were covered by PDADMAC first and decorated with CdTe nanoparticles second in order to perform energy transfer from the aggregate to the nano-particles or vice versa, depending on which size of the CdTe particles is selected. These achievements as well as the results presented here impressively demonstrate which possibilities for constructing functional units in the sense of artificial light harvesting systems are offered by these tubular J-aggregates.

## Acknowledgement

This work was supported by Deutsche Forschungsgemeinschaft (DFG) via the Max-Planck Research School on Biomimetic Systems, the Collaborative Research Group 951 ("Hybrid Inorganic/Organic Systems for Opto-Electronics (HIOS)"), and the Collaborative Research Group 448 ("Mesoscopically structured systems"). We gratefully acknowledge the Joint Lab of Structural Research between Helmholtz-Zentrum Berlin, Humboldt-Universität zu Berlin and Technische Universität Berlin. We thank Yan Qiao for support and valuable discussions and we are grateful to E. Poblenz for her help with sample preparation.

## Conflict of Interests

The authors declare that they have no conflict of interest.

## Supporting Information Available.

The contents of Supporting Information include the following: (1) Description of dynamics of aggregate formation, (2) demonstration of pH compatibility of C8S3 solutions, (3) demonstration of salt induced bundling of C8S3 aggregates, (4) zeta potential measurements of PE coated aggregates, (5) additional cryo-TEM images, (6) cryo-TEM images of C8S3/PDADMAC complexes stained with CdTe nanoparticles, (7) some useful physical data of the polycations.

# Supplementary Information

Adsorption of polyelectrolytes onto oppositely charged surface of tubular J-aggregates of a cyanine dye

*Omar Al-Khatib,*[†,‡] *Christoph Böttcher,*[§] *Hans von Berlepsch,*[§] *Katherine Herman,*[†,‡] *Sebastian Schön,*[††] *Jürgen P. Rabe,*[†,‡] *Stefan Kirstein.*[†*]

[†] Institut für Physik, Humboldt-Universität zu Berlin

[‡] IRIS Adlershof, Humboldt-Universität zu Berlin

[§] Research Center of Electron Microscopy, Freie Universtät Berlin

[††] Institut für Chemie, Technische Universität Berlin

AUTHOR EMAIL: kirstein@physik.hu-berlin.de

# Table of Contents

# 1. Optical spectra of bare C8S3 aggregates

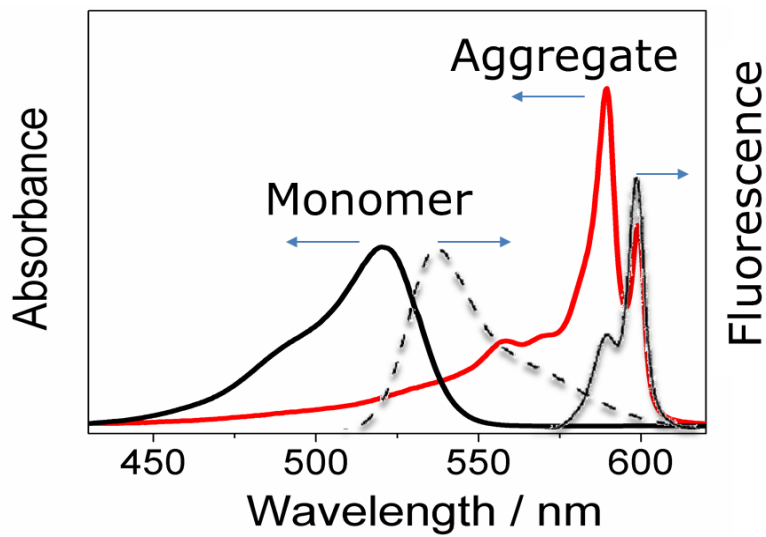

**Figure S 1: Absorbance and fluorescence of a solution of C8S3 aggregates.**

# 2. Dynamics of aggregate formation

The aggregation process begins almost immediately upon addition of the methanol stock solution to water. However, over time scales of minutes and hours the absorption spectrum still undergoes changes with increasing absorption at the J-bands. This indicates continued formation of aggregates. In Figure S 2 and Figure S 3 the change of the absorption spectra after addition of methanol stock solution to water is shown.

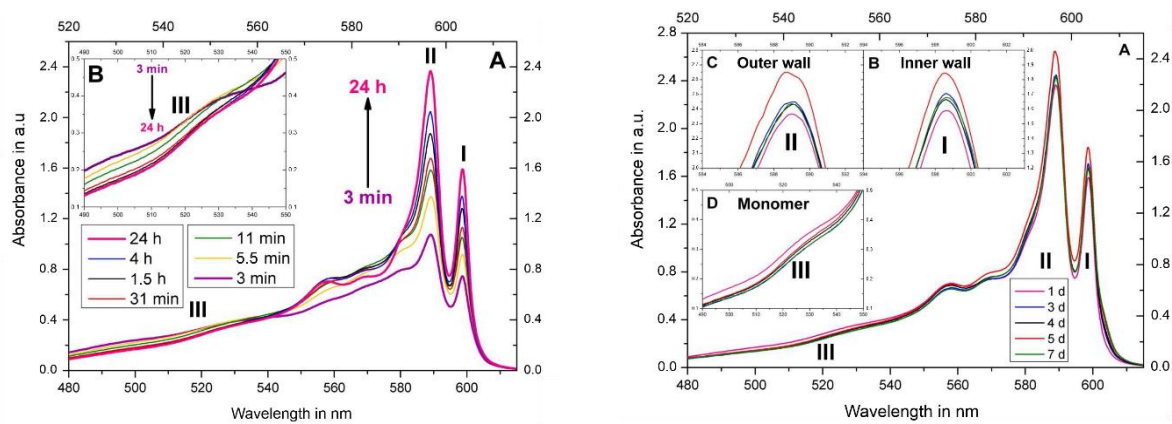

**Figure S 2: Time series of absorption spectra taken for a $6.0 \cdot 10^{-4}$ M solution of C8S3 recorded after addition of water to the methanol stock solution ($2.9 \cdot 10^{-3}$ M). The insets show magnifications of the respective areas of the total spectra.**

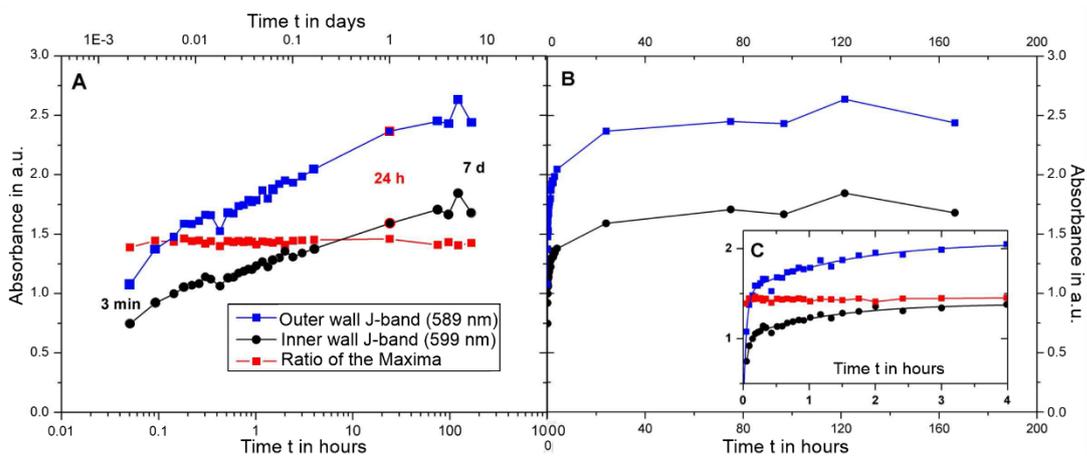

**Figure S 3:** Time evolution of the absorption of the inner wall band (599 nm) and the outer wall band (589 nm) and their ratio. (A) Plotted over log(t), (B) plotted linearly with time.

# 3. pH compatibility of C8S3 solutions

The stability of the aggregates of C8S3 with regard to solution pH was monitored by their absorption spectra. The pH was varied by addition of HCl or NaOH. For the calculation of pH value it was assumed that the pure dye solution is at neutral pH of 7. The resulting spectra are shown in Figure S 4 and summarized in Figure S 5.

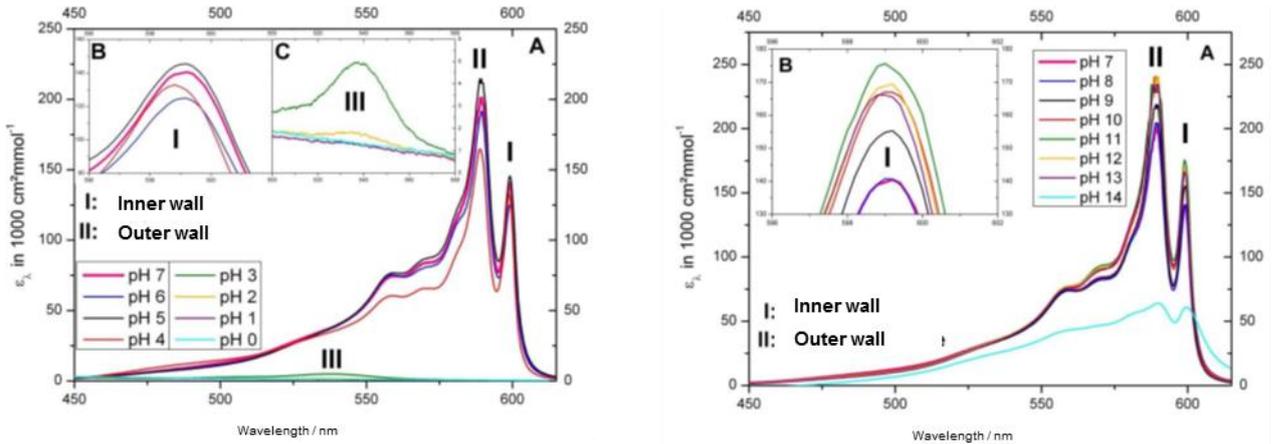

**Figure S 4: Absorption spectra of a $3.1 \cdot 10^{-4}$ M C8S3 solution at different pH values. Left: acidic range. The pH was calculated from the amount of HCl added. Right: basic range. The pH was calculated from the amount of NaOH added. (I) absorption peak of outer tube wall; (II) absorption peak of inner tube wall; (III) absorption of monomer.**

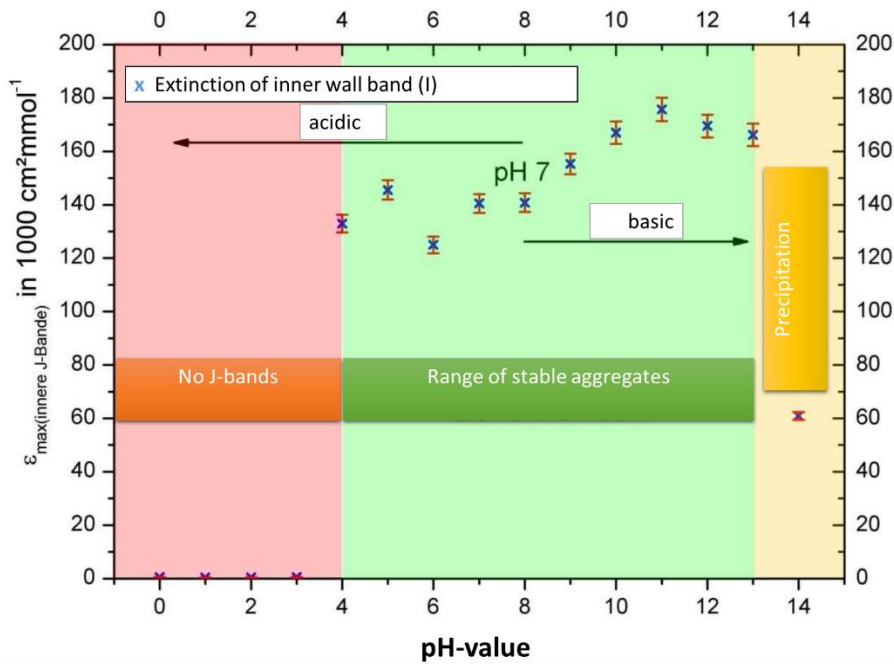

**Figure S 5: Overview of the extinction coefficients of the absorption band at 599 nm (inner wall) for various pH values.**

# 4. Salt induced bundling of C8S3 aggregates

It has been reported previously that the tubular aggregates of C8S3 show a tendency to form bundles during aging over days or weeks [1] [2]. The formation of bundled aggregates is accompanied by characteristic changes of the absorption spectrum. A typical time series measured over a period of 11 days is shown in 6a. During the bundling process the inner wall peak at 599 nm is replaced by a peak at 603 nm while the outer-wall peak decreases until it is almost completely gone at later times (not shown here, but see [1] [2]).

It seems that the process of bundling of the tubular aggregates is similar to flocculation of colloidal suspensions as described by the DLVO theory [3]. That would mean that initially the aggregates are kept at distance from each other due to electrostatic repulsion. Over time there is some probability that two aggregates overcome this electrostatic repulsion barrier by thermal motion and are attracted by short range van-der-Waals-forces. For pure C8S3 samples this process takes days or weeks.

According to the DLVO theory it should be possible to accelerate the bundling by lowering the electrostatic repulsion. This can be achieved by addition of salt since increasing ionic strength of the solvent lowers the range of electrostatic repulsion as described by the Debye screening length.

In 6b the development of the absorption spectra is shown for a solution of C8S3 with $10^2$ M NaCl solution added. This ion concentration lowers the original Debye length of 17 nm, which is due to the counter ions of the dye, by about a factor of 10. Immediately after addition of the NaCl solution the absorption spectrum looks similar to the pure dye solution after 11 days. The bundling process then develops further during 5 days and achieves a stable condition over the following days. The final spectrum is identical to the one reported in [1] after up to 3 month storage time. This enhanced bundling process due to addition of salt was confirmed also for other salts and appeared to be even more effective for two-valent salts, as predicted from DLVO theory.

It was shown by Eisele et al. [1] that the bundling essentially removes the outer layer of the tubes, resulting in a structure of packed tubes that only consist of the slightly modified inner tube, sheathed by an outer layer that encloses the entire bundle. Therefore the process of bundle formation is much more complex than described by electrostatics only. The overcoming of the electrostatic repulsion as in case of flocculation is just the first step.

Bundling of aggregates was also observed when NaCl was added to a sample of C8S3 with PAH added. In Figure S 7 the time evolution of the absorption spectra of a sample with $10^{-5}$ M PAH after addition of $10^{-2}$ M of NaCl salt is shown. Note the similarity of the spectrum after 12 h for the PAH-coated aggregates (Figure S 7a) with that after 24 h for the pure aggregates (Figure S 6b). The bundling is confirmed by cryo-TEM images, see Figure S 7b. Only very few single tube aggregates are found, the majority shows bundling. Although only the minority of the aggregates is expected to be covered by PAH this result is remarkable since it is in contrast to the previous finding for the similar dye C8O3, having carboxyl end groups instead of the sulfopropyl groups of C8S3, where a coverage with poly(vinyl alcohol) (PVA) stabilizes isolated aggregates and prevents bundling.[4]

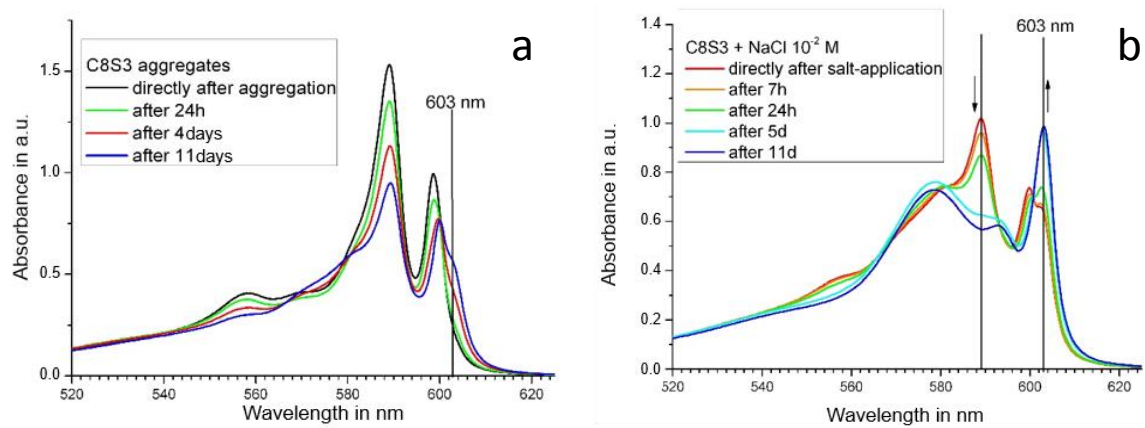

**Figure S 6:** Absorption spectra of a 3.3·10$^{-4}$ M solution of C8S3 measured at certain time steps over a period of time of 11 days. (a) Shows a pure solution of C8S3, while (b) shows a solution of same concentration but with 10$^{-2}$ M NaCl added. The new position of the inner wall peak at 603 nm that is characteristic for bundles, as well as the position of the outer wall peak at 589 nm are indicated.

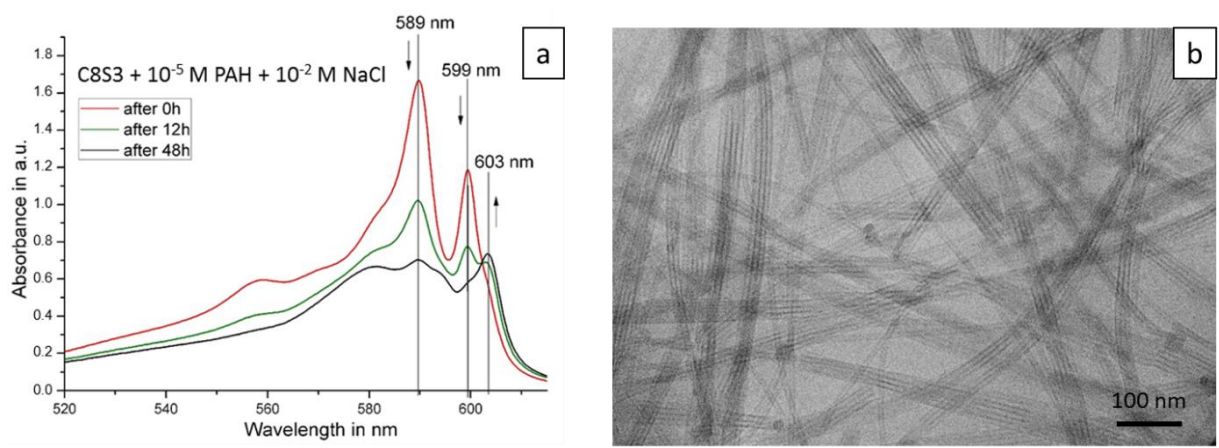

**Figure S 7:** a) Time sequence of absorption spectra of a sample of C8S3 and 10-5 M PAH recorded after addition of a 10-2 M NaCl solution. b) Cryo-TEM image of C8S3 and 10-5 M PAH with additional 10-2 M NaCl solution, taken 48 h after addition of the salt. Almost all tubular aggregates appear in bundles.

# 5. Zeta potential measurements

Measurements of the zeta-potential were carried out on a Zetasizer Nano ZS analyzer with integrated 4 mW He-Ne laser, $\lambda = 633$ nm (Malvern Instruments Ltd, U.K.). This device is able to detect particles in sizes from 0.6 up to 6 µm. Measurements were performed in phosphate buffer to determine the surface charge of the aggregates and coated aggregates. 100 ml buffer was prepared using 4.27 ml of 0.02 M $KH_3PO_4$ and 30.49 ml of 0.01 M $Na_2HPO_4$. An additional 65.24 ml of pure Milli-Q water (Millipak Express 20, 0.22 µm Flter) was added to fill the volume, resulting in a buffer pH-value of 7.6. All measurements were carried out at 25 °C using folded capillary cells (DTS 1060) in triplicate.

The zeta-potential obtained for pure C8S3 aggregates and for aggregates mixed with polycation solutions are shown in Figure S 8. For the uncoated aggregate solutions a mean zeta potential of -60 mV was found with a spread of ± 20 mV. The potential reverses sign if polycations are added and values ranging from +35 mV ± 10 mV for PEI to +62 mV ± 10 mV for PDADMAC and +68 mV ± 20 mV for PAH were found. The absolute value of the surface potentials should not be assigned any great importance since the instrument is in principle designed to detect signals from spherical particles larger than 600 nm. Here, we have cylindrical objects with very small diameter (13 nm) but length exceeding microns. The tubes deliver a sufficiently large scattering signal that allows measurement, but the conversion of the signal into a numerical value of the zeta potential is probably not correct. Also the buffer solution added to the aggregates may induce bundling that also contributes to the signal strength. However, the sign of the potential is a meaningful quantity and demonstrates charge reversal upon adsorption of the polycations.

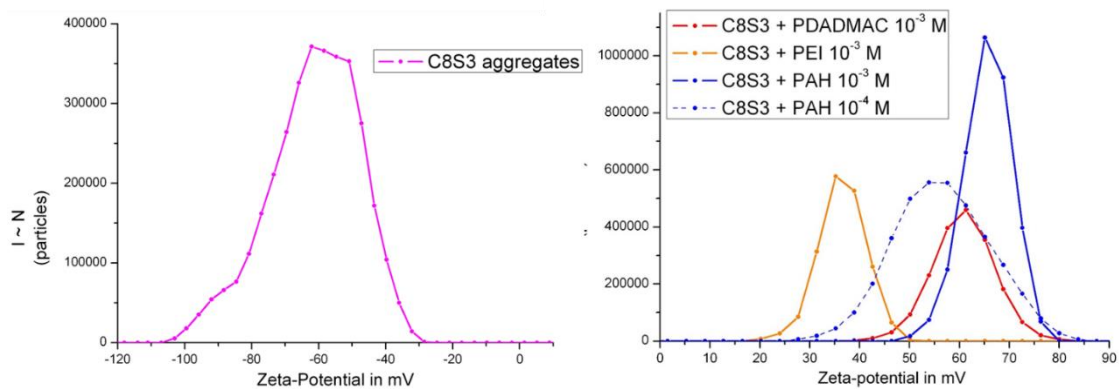

**Figure S 8: Zeta-potential plot of pure C8S3 aggregate solution and of C8S3/polycation mixtures. Each colour represents coating with a certain polyelectrolyte. All polyelectrolyte treated C8S3-solutions provide positive potentials.**

# 6. Absorption and Emission spectra of PDDA covered aggregates

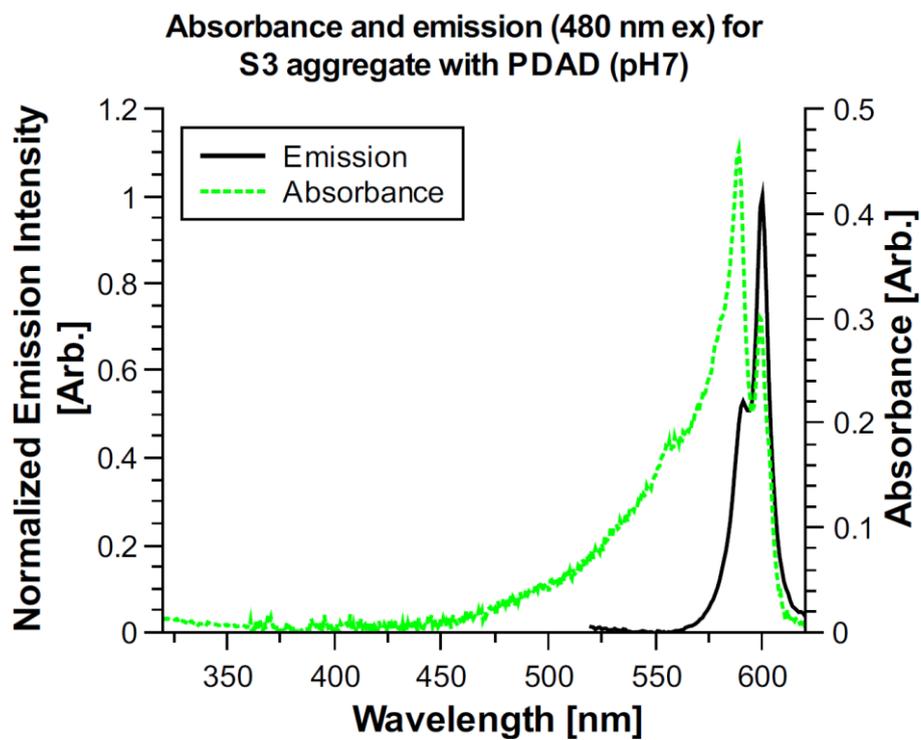

**Figure S 9:** Absorption and emission spectrum of C8S3 aggregates covered by PDADMAC. Concentration of PDADMAC was $7 \cdot 10^{-5}$ M, concentration of C8S3 was $3.3 \cdot 10^{-4}$ M.

# 7. Additional cryo-TEM images

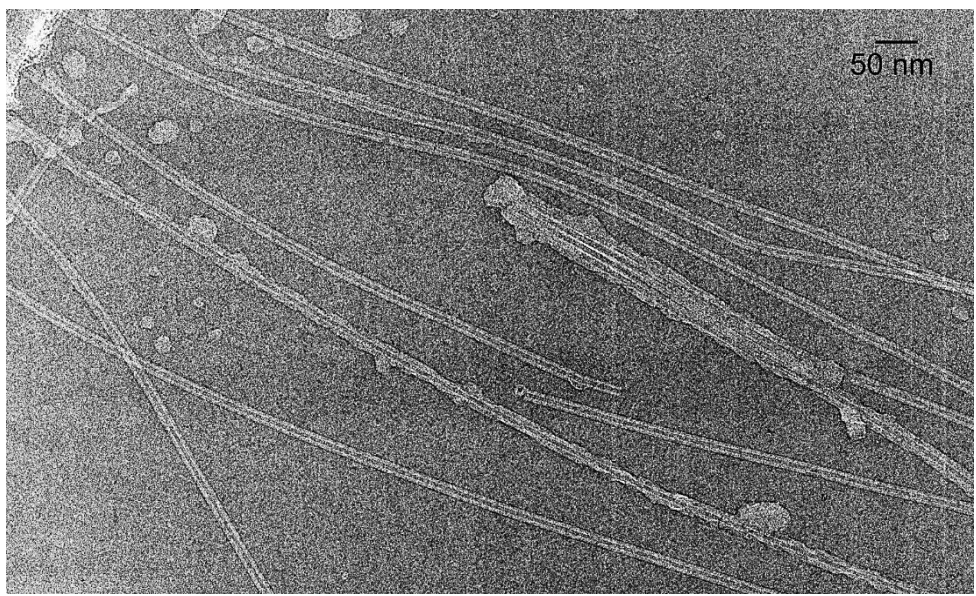

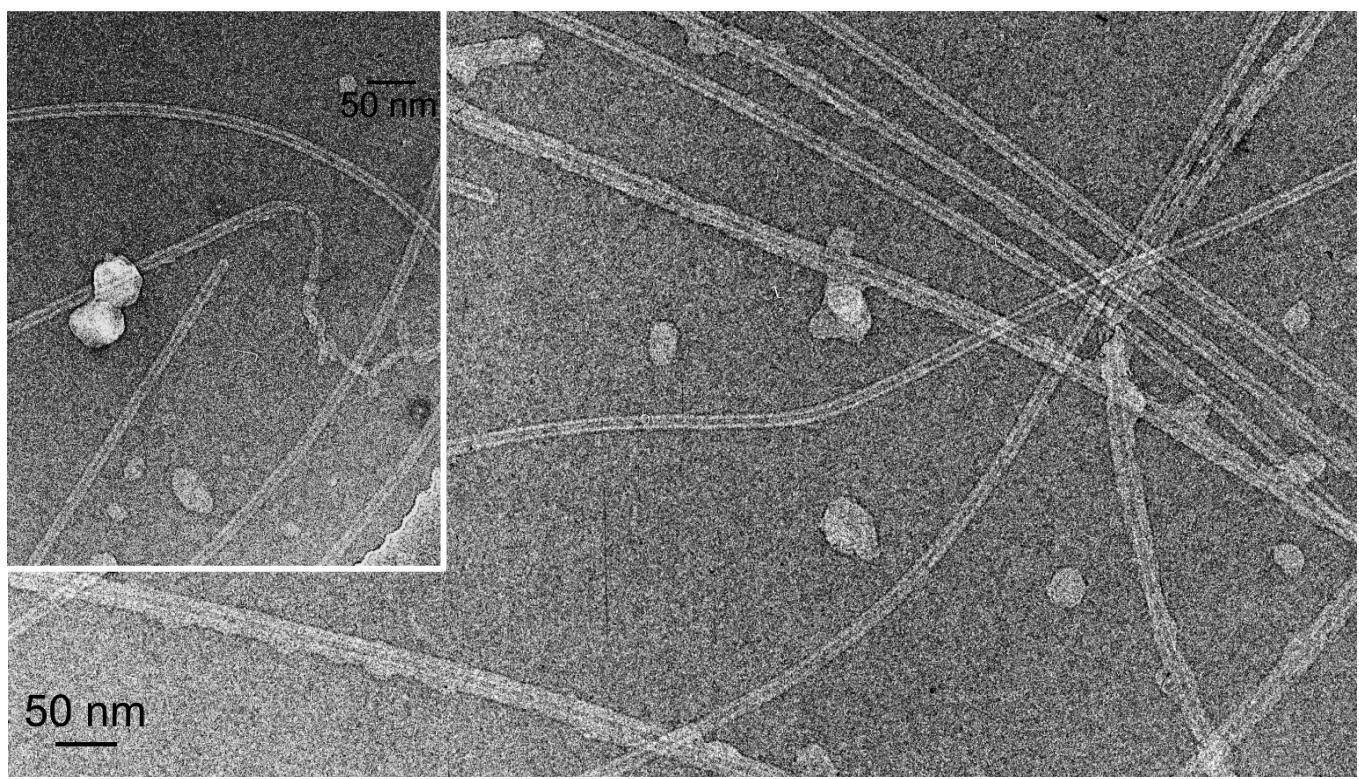

**Figure S10 (a) and (b):** Overview images of $3.3 \cdot 10^{-4}$ M C8S3 aggregate solution after addition of $10^{-5}$ M solution of PAH in equal volume. The insert of (b) shows part of an aggregate where the tubular structure is destroyed. Its typical width nevertheless has roughly the size of the original aggregates. We call this type of structures "distorted aggregates".

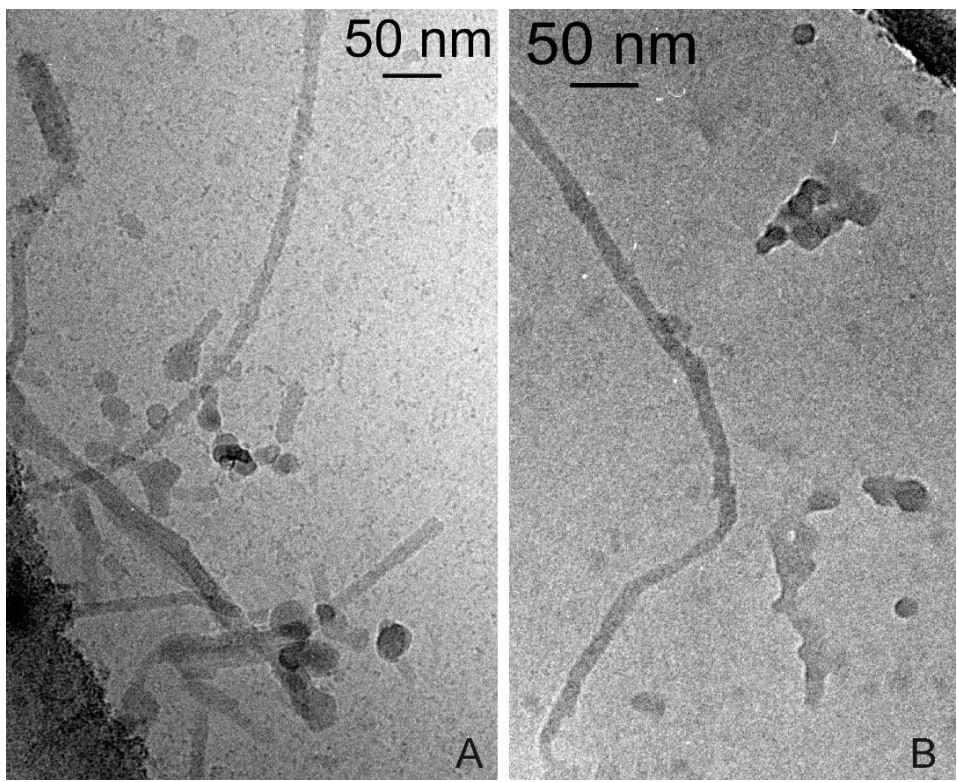

**Figure S 11: Cryo-TEM images of 3.3·10⁻⁴ M C8S3 aggregate solution with 10⁻³ M solution of PAH added in equal volume. The tubular structure is no longer visible and has a similar appearance as shown in Figure S10b (insert).**

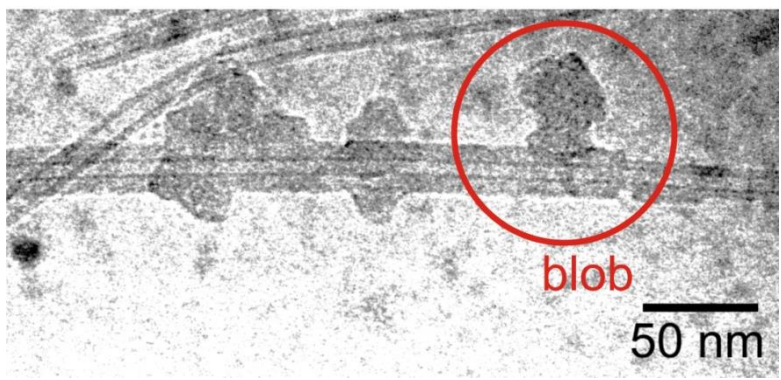

**Figure S 12: Cryo-TEM images of 3.3·10⁻⁴ M C8S3 solution after addition of 10⁻⁴ M of low molecular weight PDADMAC (PDA100) solution. Pure aggregates (top of image) as well as aggregates decorated by blob-like structures and an additional double layer of dye are visible.**

# 8. Staining of C8S3/PDADMAC complexes with CdTe nanoparticles

In a previous publication by Y. Qiao et al. [5] aggregates covered by PDADMAC were stained with CdTe nanoparticles. This was done by adding 8 uL of 1 mM PDA100 (average MW 100 000 – 200 000, Sigma-Aldrich) to 100 uL of C8S3 J-aggregate solution to obtain a final PDADMAC concentration of approximately $10^{-4}$ M. . The solution was kept in dark for 3 hours before adding 20 μL MPA-capped CdTe (PlasmaChem, Germany) QDs solution (0.5 mg/mL). The diameter of the CdTe nanoparticles is 2 – 3 nm. The final solution was stored in the dark for another 3 hours before measurements were performed. The MPA-capped CdTe particles are negatively charged and therefore do not adsorb directly on the negatively charged dye aggregate. In Figure S 13a an aggregate is shown where isolated PDADMAC chains decorated with CdTe particles are visible wrapping around the tubular aggregate. It is noticeable that no CdTe particles have adsorbed at the section with increased wall thickness. This was found repeatedly in various images. Hence, one may conclude that this area is still negatively charged and has no PDADMAC coverage. In Figure S 13b an aggregate is shown that is almost homogeneously covered with CdTe particles. Additionally, blobs are seen that are also covered by the particles. This implies that the surface charge of this aggregate was completely reversed by a layer of PDADMAC. Also the blobs are clearly identified as positively charged. The average diameter of the CdTe covered aggregate[5] of 20.4±0.5 nm fits well to the outer diameter of the bare aggregate (13±0.5 nm) plus two times the diameter of the nanoparticles (3±0.5 nm). The thickness of the PDADMAC layer between aggregate and nanoparticles then has to be in the range of 1±0.5 nm.

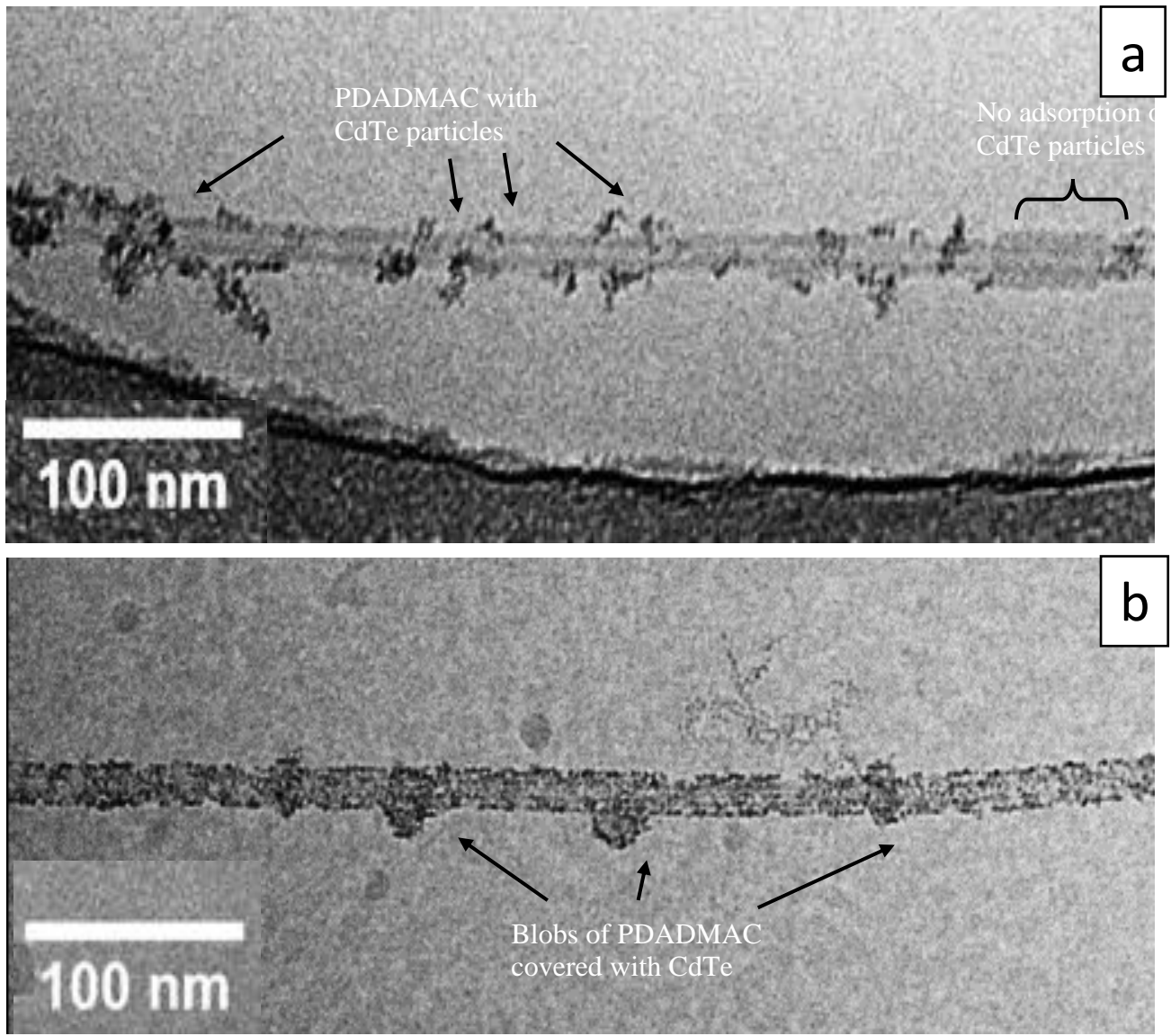

**Figure S 13: Cryo-TEM images of C8S3 tubular aggregates covered with PDADMAC and stained with CdTe nanoparticles. (a) Aggregate wrapped by sporadic PDADMAC chains decorated with CdTe nanoparticles. The thickened section doesn't show any adsorbed CdTe particles (images are courtesy of Y. Qiao and were already published in similar form in [5]).**

## 9. Useful physical data of the polycations used

|  | PDAD100 | PDAD400 | PAH | PEI |
|---|---|---|---|---|
| Average molecular weight $M_w$ g/mol | 100000 | 400000 | 17000 | 60000 |
| Average number of monomers N | 620 | 2480 | 180 | 1360 |
| Molecular weight of monomer / g/mol | 161.6 | 161.6 | 93.5 | 44.1 – 472.7 |
| Coulomb length $l_c = 1/\tau$ (α=1, fully charged) / nm | 0.5 | 0.5 | 0.22 [6] (1.25 for 20% charged) | 0.38 |
| Line charge density τ (α=1, fully charged) / e/Å | 0.2 | 0.2 | 0.4 | 0.26 |
| $pK_a$ (pH for α=0.5) |  |  | 8.9 [7] | 8.2 [8] |
| Radius of gyration $R_g$ / nm | 22 [9] | 40 [9] | 4.8 … 6 [6] (pH=6 … 9), $10^{-2}$ M salt | 22 [10] 1 M salt |